# SLIM: Simultaneous Logic-in-Memory Computing Exploiting Bilayer Analog OxRAM Devices

Sandeep Kaur Kingra[1], Vivek Parmar[1], Che-Chia Chang[2], Boris Hudec[2], Tuo-Hung Hou[2], and Manan Suri[1*]

[1]Department of Electrical Engineering, Indian Institute of Technology-Delhi, Hauz Khas, New Delhi, 110016, India.

[2]Department of Electronics Engineering and Institute of Electronics, National Chiao Tung University, Hsinchu, 300, Taiwan.

[*]manansuri@iitd.ac.in

**Von Neumann architecture based computers isolate/physically separate computation and storage units i.e. data is shuttled between computation unit (processor) and memory unit to realize logic/ arithmetic and storage functions. This to-and-fro movement of data leads to a fundamental limitation of modern computers, known as the *memory wall*. Logic in-Memory (LIM) approaches aim to address this bottleneck by computing inside the memory units and thereby eliminating the energy-intensive and time-consuming data movement. However, most LIM approaches reported in literature are not truly “simultaneous” as during LIM operation the bitcell can be used only as a Memory cell or only as a Logic cell. The bitcell is not capable of storing both the Memory/Logic outputs simultaneously. Here, we propose a novel ‘Simultaneous Logic in-Memory’ (SLIM) methodology that allows to implement both Memory and Logic operations simultaneously on the same bitcell in a non-destructive manner without losing the previously stored Memory state. Through extensive experiments we demonstrate the SLIM methodology using non-filamentary bilayer analog OxRAM devices with NMOS transistors (2T-1R bitcell). Detailed programming scheme, array level implementation and controller architecture are also proposed. Furthermore, to study the impact of introducing SLIM array in the memory hierarchy, a simple image processing application (edge detection) is also investigated. It has been estimated that by performing all computations inside the SLIM array, the total Energy Delay Product (EDP) reduces by ~ 40x in comparison to a modern-day computer. EDP saving owing to reduction in data transfer between CPU ↔ Memory is observed to be ~ 780x.**

Over the past few decades, the performance gap between the computing unit (where the data is processed) and the memory unit (where the data is stored) is increasing, popularly known as the memory wall[1]. It is observed that for many computing tasks, most of the time and energy is consumed in data transfer between the processing unit and memory unit, rather than the computation[2]. To tackle this bottleneck, various solutions have been proposed, targeting from component level to the system architecture level. Measures include the extensive use of spatial architectures (distributed on-chip memory that is closer to the computation unit) enabling parallelism using vector processing unit, with large number of cores[3]. Furthermore, accelerators have been designed to match the exact data flow for specific computing algorithms[4]. Three-dimensional memories, commercialized as hybrid memory cube[5] and high bandwidth memory[6] chips have been proposed to meet the requirements of high data-transfer rate and high memory density. These deliver an order of magnitude higher bandwidth and reduce access energy by up to 5x relative to existing 2-dimensional DRAMs[7]. Moving further, emerging non-volatile memories (NVM) have been introduced into the traditional memory hierarchy to minimize the ‘gap’ between computing and the data units[8]. However, the most promising solution to eliminate such bottlenecks is governed by logic in-memory (LIM)/ processing in-memory (PIM) approach, where computations are carried out *in situ*, exactly where the data is located[9].

LIM offers a clear advantage by totally removing the latency and energy burdens of the memory wall. However, this new architecture requires computational memory devices that can both store data and compute at the same time, usually by device physics or other physical laws. Due to their unique properties, emerging NVMs have proved themselves as promising technologies for LIM approaches. NVMs offer non-volatility along with a large resistance window supporting clear discrimination of states ‘1’ and ‘0’ (few offer multilevel capability (MLC) as well).

Recent LIM approaches[10-15] focus on realizing novel logic gate concepts with lower energy consumption and area footprint. These approaches carry out digital Boolean operations with difference in the type of input, the type of output and the physical operation to describe the logic function. Using memristive bipolar resistive switches (BRS) and complementary resistive switches (CRS), multiple Boolean functions have been realized using three sequential steps[10]. Using three memristors, multiple Boolean functions (OR, AND, NOR, NAND, NOT) have been realized using two steps in MAGIC[14]. But these implementations had sneak paths due to the absence of selectors along with memristors. To mitigate this issue, a logic methodology using 1T-1R structure to implement functionally complete Boolean logics has also been proposed[15]. However, this methodology demands two steps (initializing and writing) to implement any logic gate. The aforementioned approaches[10-15] have considerably increased the application scope of emerging NVM devices by extending them to beyond storage computational functions. However none of the LIM approaches discussed in literature has shown truly 'simultaneous' Logic-inside memory. While the existing LIM approaches offer ways to use memory cells for Logic operations, one must realize that whenever a particular memory cell is performing a Logic-operation it cannot be used for its inherent memory or storage functionality simultaneously. Fundamental or ground-breaking impact of In-Memory Computing can only be realized if the same memory bitcell can be used for both Storage-function and Logic-function simultaneously in space (silicon area) and time (clock cycles).

In this work, we propose the first ever true- 'Simultaneous Logic-in-Memory' (SLIM) methodology using non-filamentary analog OxRAM devices and NMOS transistors. We propose a 2T-1R SLIM bitcell and a novel array-level programming methodology. Figure 1(a) shows three different architectures: von Neumann architecture that is used nowadays in modern computers, recent LIM architecture and the proposed SLIM architecture. We show that using our SLIM approach with analog OxRAM it is possible to simultaneously- (i) perform Logic operation in-situ, (ii) store output of the Logic operation, (iii) preserve the previously stored Memory state (i.e. value stored on the cell prior to the Logic operation) and (iv) read both final Logic output and stored memory values, all while using the same bitcell.

**Results:**

**SLIM Bitcell:**

Figure 1(b) shows our proposed 2T-1R SLIM bitcell. Analog bilayer OxRAM device (Ni/3 nm $HfO_2$/7 nm Al-doped-$TiO_2$ (ATO)/TiN stack) is integrated with 2 NMOS transistors. The bottom electrode (BE, $V_{out}$) of OxRAM is connected to NMOS drain terminals, and the top electrode (TE, $V_1$) is used for applying direct programming signals (P1/P2). OxRAM device exhibiting analog resistive switching characteristics was fabricated using CMOS compatible process with a 50 µm x 50 µm active device area, where the ATO as well as $HfO_2$ were deposited using plasma-enhanced atomic layer deposition (PE-ALD). Transmission electron microscopy (TEM) cross-section image of the device stack is shown in Figure 1(c), where amorphous dielectric bi-layer is seen deposited on the TiN BE film with high uniformity, a fingerprint of ALD. The fabrication flow is described in Methods section. Figure 1(d) shows measured DC IV curves of the OxRAM device under study (see supplementary Figure 1 for impact of C2C variability on DC IV curves). Owing to its non-filamentary, analog resistive switching, the stack exhibits highly reproducible behavior where the resistance can be gradually tuned in a continuous manner by an applied voltage stimuli, as shown in supplementary Figure 2. High reproducibility of the electrical characteristics with small variations can be attributed to the high uniformity of dielectric films deposited by PE-ALD and the non-filamentary switching nature of these OxRAM devices. The gate terminals of both NMOS transistor ($V_{G1}$, $V_{G2}$) are used to load operands as control signals during Logic operations. To physically realize the SLIM bitcell, discrete CMOS devices [CD4007UB] were connected to OxRAM devices through wire-bonding and packaging (experimental setup is shown in supplementary sheet (supplementary Figure 3, 4) along with the $I_D$-$V_{DS}$ and $I_D$-$V_{GS}$ characteristics of the NMOS transistor). NMOS transistors help with: (i) SLIM realization, (ii) current-compliance, and (iii) sneak path minimization. The gate voltage ($V_G$) decides the transistor resistance ($R_{NMOS}$) and current through the device. In 2T-1R bitcell, for OxRAM device SET, we use $V_G$ = 4 V, corresponding to $R_{NMOS} \approx 1$ kΩ and for RESET we use $V_G$ = 10 V. OxRAM device fabricated for this study is for demonstration purpose, further scaled OxRAM devices with integrated on-chip NMOS transistors (or selectors) will lead to a significant reduction in operating voltages.

**Concept of SLIM and SLIM-State Definitions:**

SLIM fundamentally differs from other LIM architectures previously reported in literature[10-15]; as it allows the use of a given memory cell, to simultaneously perform a Logic operation while retaining its actual Memory state. To achieve simultaneous non-destructive Logic operation we propose the following methodology- First, we define 4 distinct resistance states (labelled: '11', '10', '01' and '00' in Figure 2(a)) among the continuum of attainable OxRAM resistance levels (supplementary Figure 2). The distinct states are selected on basis of: (i) programming reproducibility, (ii) ease of inter-state transitions between the 4 levels; i.e. requirement of fewer types of programming pulses (signals P1, P2, P3 in Figure 2(c)) and (iii) compatibility for a simultaneous Memory/Logic- Read operation. Each of the 4 selected states ('11', '10',

‘01’, ‘00’) are assigned both- Logic (‘1’/‘0’) and Memory (LRS/HRS) definitions respectively. The sense-amplifier threshold and Memory state sensing window is defined such that 2 states lie in memory LRS sense region while other 2 in memory HRS sense region respectively. In Figure 2(a), states ‘11’ and ‘10’ lie in memory LRS region and are assigned logic ‘1’ and logic ‘0’ values respectively. While states ‘01’ and ‘00’ lie in memory HRS region and are assigned logic ‘1’ and logic ‘0’ values respectively. Thus, with such state assignment, the system has two representations each for; logic ‘1’, logic ‘0’, memory HRS and memory LRS. While executing any Logic operation, the SLIM programming scheme permits state transitions only within the logic (‘1’/‘0’) levels of a particular memory sense region (i.e. logic ‘1’↔ ‘0’ within HRS or within LRS sense regions are permitted, but logic ‘1’ ↔ ‘0’ through HRS ↔ LRS is not permissible). Thus any initially stored Memory state on the bitcell can be preserved even after executing a Logic operation. A single read-operation can simultaneously read and decode both Memory- and Logic-values on the device. Figure 2(b) shows experimentally programmed distributions of the 4 SLIM resistance levels and their cycling endurance. Figure 2(c) illustrates the required programming signals (P1, P2, P3) and possible SLIM inter-state transitions while implementing any Logic or Memory operation. On application of P1/P2 pulses, the OxRAM device's resistance is gradually lowered while with application of P3 pulse the resistance is increased.

**Experimental validation of Memory-Write Operation:**
These refer to purely Storage operations, with the purpose of just storing a bit (‘1’/‘0’) on the 2T-1R bitcell. Data can be stored on the bitcell through Memory Write- ‘1’ (i.e. absolute Memory state ‘11’, LRS) or Memory Write- ‘0’ (i.e. absolute Memory state ‘01’, HRS). During Memory Write operation, the bitcell is programmed to absolute Memory states (as shown in Figure 2(a)). Figure 2(d) shows precise bitcell programming methodology for the Memory Write operations. Supplementary Figure 5 shows detailed experimental results for Memory Write ‘1’ (supplementary Figure 5(a-c)) and Memory Write ‘0’ (supplementary Figure 5(d-f)) from all possible initial states.

**Experimental validation of 2T-1R SLIM Logic operation (Universal Gate: NOR gate) and SLIM protocol:**
These refer to pure Logic operations to be performed on the 2T-1R SLIM bitcell (ex- NOR, AND, NOT etc.). Table 1 presents the SLIM state-mapping and truth-table for realizing a 2-input NOR gate operation using a single 2T-1R SLIM bitcell. Prior to executing Logic operation the bitcell may initially contain any stored Memory state (i.e. absolute Memory state ‘11’ or ‘01’), as a consequence of a preceding Memory operation(s). While realizing Logic operation, SLIM bitcell may undergo a state change but the overall Memory state is preserved. Table 1 shows that previous Memory state is effectively preserved after each NOR Logic operation, by the virtue of SLIM state assignment and programming methodology described in Figure 2(a, c). The circuit schematic and NOR programming signals for all possible 1-bit, 2-input (a, b) operand combinations are shown in supplementary Figure 6. Operands *a/b* are mapped to $V_{G1}/V_{G2}$ respectively, while signal P3 is applied on terminal $V_2$, keeping $V_1$ grounded. When either *a* or *b* is ‘0’, the corresponding NMOS is OFF. However, when either *a* or *b* is ‘1’ (logic high), the corresponding $R_{NMOS}$ < 1 kΩ ($V_G$ = 10 V), as a result the applied programming signal drops across OxRAM and it undergoes RESET switching. Figure 3(a)(i-iv) show experimental implementation of NOR-Logic operation on SLIM bitcell for initial device state: ‘11’ (i.e. stored Memory LRS/’1’), and Figure 3(a)(v-viii) for initial device state: ‘01’ (i.e. stored Memory HRS/‘0’). It is evident that for two consecutive identical RESET pulses (P3), owing to non-filamentary switching, the OxRAM resistance gradually increases and the current flowing through the device decreases. Similar OxRAM behaviour in response to consecutive RESET pulses (P3) was also observed while performing SLIM Memory operations (supplementary Figure 5(d, e)). All experiments repeatedly validate preservation of initial Memory state post Logic operation. This is first of a kind simultaneous implementation reported in LIM literature.

**Array-level Implementation of SLIM, realization of complex logic and Boolean-functions beyond NOR:**
To avoid possible resistance state saturation during multiple consecutive SLIM Logic ↔ Memory operations inside a large 2T-1R SLIM array, we define an intelligent SLIM control protocol with an added Refresh block (Figure 3(b), 4(a)). In Refresh scheme, initial state of the bitcell is read. If it is in absolute Memory state (i.e. ‘11’ or ‘01’), logic implementation can be performed directly by following programming scheme defined in Figure 2(c) and Table 1. However, in case the state of bitcell is in non-absolute Memory state (i.e. ‘10’ or ‘00’), *Refresh* is applied (i.e. signal P2 at $V_1$, $V_2$ = gnd, $V_{G1}$=$V_{G2}$= 4 V). As shown in Figure 2(c), application of P2 always restores the bitcell from a Logic state ('10' or '00') to its corresponding absolute Memory state ('11' or '01') respectively. This enables the same bitcell for another consecutive Logic operation while effectively preserving the previous Memory state. The Refresh scheme can be further optimized by only performing it periodically at the array level, using Tag-bytes to track the last operation of the SLIM Matrix (Mat) unit (refer supplementary Figure 7 for further details). SLIM matrix indicates 2-D SLIM bitcell array. Once, entire Mat has been used for Logic operations, row-wise Refresh operation may be enabled.

Figure 4(a) presents the SLIM controller implementation encompassing all details of Figure 2(b-d). It has three blocks to control Memory Write-, Logic- and Read-operations. There is an additional control signal to choose either Memory/ Logic mode. Within the Logic block, an additional intelligence is built to implement refresh mechanism. Figure 4(b) shows the implementation of a 1-bit half adder (HA) using the proposed SLIM NOR logic with accurate cycle and bitcell location mapping. Table 2 illustrates the extension of SLIM methodology to Logic operations beyond NOR (while using the same programming signals). It should be noted that in all these cases only one SLIM bitcell is required to realize the Boolean function.

**Performance Analysis:**

It is evident that SLIM bitcell is logically complete as it is possible to realize NOR gate (i.e. a universal gate). All other basic gates and arithmetic functions can be realized by mapping the desired function using NOR gate. To understand the impact of proposed SLIM methodology for logic computation we have derived energy and latency estimations for few basic logic gates normalized w.r.t. the values for NOR gate (i.e. a single 2T-1R SLIM bitcell). Results have been summarized in Table 3. OxRAM devices used in this study have larger area, thereby it exhibits higher delay and energy costs. To do a fair comparison with state of the art architectures, we have used device parameters of advanced bilayer filamentary $HfO_x$ devices[16, 17] (listed in supplementary Table 1). Figure 5(a) highlights the performance comparison for 64-bit Logic operations performed using a conventional CPU (Intel Core i5-2500 Sandy Bridge CPU) and SLIM bitcell assuming data is to be fetched from DRAM/SLIM array. For all given Logic operations, we can observe a minimum EDP benefit of ≈ 4x. To further study the impact of introducing SLIM array in the memory hierarchy on a real-world workload, a simple image processing application of edge detection is analyzed. Edge detection is mainly based on convolution of an image with a fixed filter. An image of size 64x64 (shown in Figure 5(b)) and a 3x3 'Sobel filter' are used for performing edge detection. The convolution operation here involves 9 multiply and 9 add operations. To minimize the complexity of implementation, 4-bit precision for operation is considered using SLIM array in comparison to the full 8-bit precision used by CPU. Carry-save architecture is used for implementing 4-bit multiplier. Furthermore, 4-bit ripple carry adder is used for calculating the final result. For SLIM array, performance and energy values mentioned in supplementary Table 1 are selected whereas for CPU, performance values from technical reports[18,19] are considered. Figure 5(b) presents the results for edge detection using high precision CPU and low precision SLIM array respectively. Table 4 summarizes the relative EDP comparison in terms of data transfer and computation operations for both approaches. It is assumed that data (for computation) is fetched from the main memory (DRAM array) for the CPU case. Furthermore, the energy and latency penalties arising due to cache misses are also taken into account. Instructions for SLIM Logic operation are incorporated within the Memory instructions to minimize the overhead. By performing all computations inside the SLIM array, the total EDP for the application reduces up to 40x (summarized in Table 4). This performance benefit will be more pronounced with further improvements in device technology and compiler based optimizations. It is also observed from Table 4 that data transfer savings in EDP is ~ 780x due to reduction in data transfer (CPU ↔ Memory) as a result of in-memory computation. For computation using in conventional CPU, all operands need to be fetched from main memory (DRAM). However for doing computation using SLIM, operands are already present in SLIM array (as it resides at the same position as DRAM). Due to this, only results need to be fetched by the CPU. NVM devices such OxRAM are more power-efficient in terms of read energy and comparable in terms of latency when compared to DRAM. With reduction in data transfer between CPU ↔ Memory, huge savings in EDP is observed. Table 5 qualitatively benchmarks our SLIM methodology w.r.t. other leading LIM approaches published in literature[10-15]. SLIM helps in realizing the Logic operation in single cycle when compared to other methodologies and offers simultaneous storage of Memory and Logic outputs.

**Discussions**

We present the first ever true 'Simultaneous Logic-in-Memory' methodology (SLIM) to overcome the limitations of conventional LIM systems. While all other approaches in literature succeed at using a memory cell for logic, however when the cell is indeed used for logic it's no longer used for memory at that time, and it also loses its initial stored state. Whereas in our proposed approach we show for the first time that true LIM should not implement Logic and Memory at the cost of each other. SLIM approach is similar to the computing mechanism in the human brain, where information is processed in sparse networks of neurons and synapses, without any physical separation between computation and memory[20]. Synapses and neurons also perform computation while storing state information i.e. they perform both computation and storage functions. The proposed SLIM approach has a remarkable potential in terms of ultimate 'effective-area' reduction for LIM systems. We experimentally validate NOR logic on 2T-1R SLIM bitcell arrays using bilayer analog OxRAM and NMOS devices. Realization of multiple complex logic functions are shown. Detailed programming/refresh schemes, array-level implementation and controller architecture are also proposed. If incorporated in future MLC SSDs, the latent storage media can act as free computational resource without disturbing the user's stored data; thus enabling the SSD to store and compute at the same time.

## Methods

**Device Fabrication:**

Analog resistive switching OxRAM stacks of Ni/3 nm $HfO_2$/7 nm Al-doped-$TiO_2$ (ATO)/TiN (top to bottom) structure were fabricated by following a CMOS compatible process. First, 100 nm thick TiN BE film was deposited on thermal-$SiO_2$ (500 nm)/Si wafer by physical vapor deposition (PVD), RF magnetron sputtering. The BEs were then patterned by optical photolithography (first mask) and dry etching using inductively-coupled plasma (ICP). The bottom, 7 nm thick ATO dielectric, was then deposited by interchanging varying amount of $TiO_2$ and $Al_2O_3$ PE-ALD cycles, using TDMATi (Tetrakis(dimethylamido)titanium) and TMA (trimethylaluminum) as metal-organic precursors and $O_2$ plasma as a reactant. Upper, 3 nm thick dielectric $HfO_2$ film, was deposited using TDMAHf (Tetrakis(dimethylamido)hafnium) and $O_2$ plasma. All depositions were carried out at 250 °C using Veeco-CNT Fiji F202 remote plasma hot-wall reactor PE-ALD system. The TE pattern (similar to the BE pattern but rotated 90°) was defined using second optical photolithography mask and 100 nm thick Ni top electrode film was deposited by DC sputtering and patterned using lift-off technique. This way, a 12 x 12 crossbar array was formed with 50 µm wide TiN and Ni perpendicular electrode lines sandwiching the dielectric bilayer, forming an OxRAM device with 50 µm x 50 µm active area at each crosspoint. Final photolithography (third mask) and ICP dry etching step was performed to open the contact windows (etch the dielectrics) to the BE contact pads. Wire bonding and packaging were the final steps for the OxRAM encapsulation. Electrical characterization was performed using Keithley 4200 SCS with 4225 PMU unit.

**Electrical measurements for SLIM bitcell characterization:**

We used Keithley 4200 SCS parameter analyzer to make electrical measurements for results shown in Figures 1(d), 2(a, b), 3(a) and supplementary Figures 1, 2, 3 and 5. For Figure 1(d), dc signal was applied across the bilayer OxRAM device, using Keithley 4210 high power SMU and the current was measured using the other SMU channel. For supplementary Figure 2, we applied voltage pulses across the OxRAM device using Keithley 4225 PMU. Consecutive SET/RESET pulses with amplitude (3 V/ -5.5 V)/(3 V/ -3 V) were applied. After each SET/ RESET pulse, a read signal, $V_{READ}$ = -0.4/ -1.5 V was applied. The current through the device was measured and corresponding conductance value was calculated. NMOS transistors from CD4007UB CMOS dual complementary pair plus inverter IC chip were used. As shown in supplementary Figure 3, the characteristics of NMOS transistor were measured using Keithley 4210 high power SMU. Gate voltage was supplied from external power supply and drain voltage was applied using Keithley 4210 high power SMU. While characterizing NMOS transistors, a passive resistor of 1 kΩ was placed in series.

An in-house customized measurement system was developed to implement 2T-1R bitcell. As shown in supplementary Figure 4, a chip of standalone OxRAM devices was wire bonded with the NMOS transistors to realize a complete 2T-1R bitcell. For Figure 2(a) different pulse signal (i.e. P1/P2/P3) were applied using Keithley 4225 PMUs. Once the resistance levels for realizing four different states were locked, Memory- and Logic operations were realized using multiple Keithley 4225 PMU units (shown in Figure 3(a) and supplementary Figure 5).

**Endurance test:**

The endurance test was conducted using constant-voltage-stress (CVS) program. Four SLIM states selected for this study were achieved using programming signals (P1/P2/P3) in one sequence, and the sequence was repeated by 200 times to test the endurance.

**Performance Analysis:**

For calculating, the performance estimates for SLIM bitcell, we used an approximate model for deriving switching energy (equation provided below).

$$Energy\ (SLIM\ op) = \frac{\sum_i^N Switch_events(SLIM\ op, x_i)}{N}$$

where N denotes total possible input combinations of input $x_i$. Switch_events function is a mathematical model that determines state transition events at each node assuming NOR based logic for computing the specified Logic operation. The energy for single bit logic gates has been derived and extrapolated for multiple-bits (assuming multi-bit operations can be realized as parallel computation using multiple banks in the SLIM bitcell array). Latency is derived by estimating total number of levels required for computing the logic function based on the computation graph derived using NOR logic. Edge detection application used as case study, involves convolution of 64x64 image with a 3x3 sized filter. Sobel edge detection operator has been used as the filter. The operation involves 9 x MUL and 9 x ADD operations of 4-bit precision. The energy estimate is based on average switching energy for all operations. SLIM array sized 4 kB with MAT size of 8 bytes spread across 16 banks with 32 MATs/bank. Total parallel single bit SLIM ops possible would be 32x16x8 = 4096. The ops are of 4-bit precision and at max 8 parallel ops are performed simultaneously in the pipeline, therefore maximum operations that can be performed in parallel are 128. Using this we derived the worst case latency and EDP values for the application. For the (CPU+DRAM) case, CPU used for comparison is Intel Core i5-2500 running at 3.3 GHz. For each Sobel operation, the CPU uses 9 x IMUL, 9 x ADD, 18 x LOAD and 9 x STORE operations. Cycle latency and average energy dissipation for each operation is used for calculating EDP. Total Sobel operations performed are 64x64=4096. For data transfer, the bus-width used for all memory interfaces is 128-bit.

**Data availability:**

The data that support the plots within this paper and other findings of this study are available from the corresponding author upon reasonable request.

## Acknowledgements

This work was partially supported by the Department of Science & Technology (DST), SERB-EMR, MHRD-Imprint (RP03417G), Government of India, IIT-D FIRP grants, Ministry of Science and Technology, Taiwan, and Ministry of Education, Taiwan. Authors are thankful to Prof. Wen-Wei Wu's group from Dept. of Materials Science and Engineering of NCTU for providing help with the TEM characterization.

## Contributions

M.S. and S.K.K. conceptualized the original SLIM methodology and initiated the work. T.H.H, C.C.C and B.H. fabricated the OxRAM devices and carried out device level characterization. SLIM bitcell characterization and experiments were done by S.K.K. Bitcell performance analysis was performed by V.P. M.S. and S.K.K. wrote the manuscript. All discussed results and reviewed the manuscript.

## Competing interests

The authors declare no competing financial interests.

# Figures:

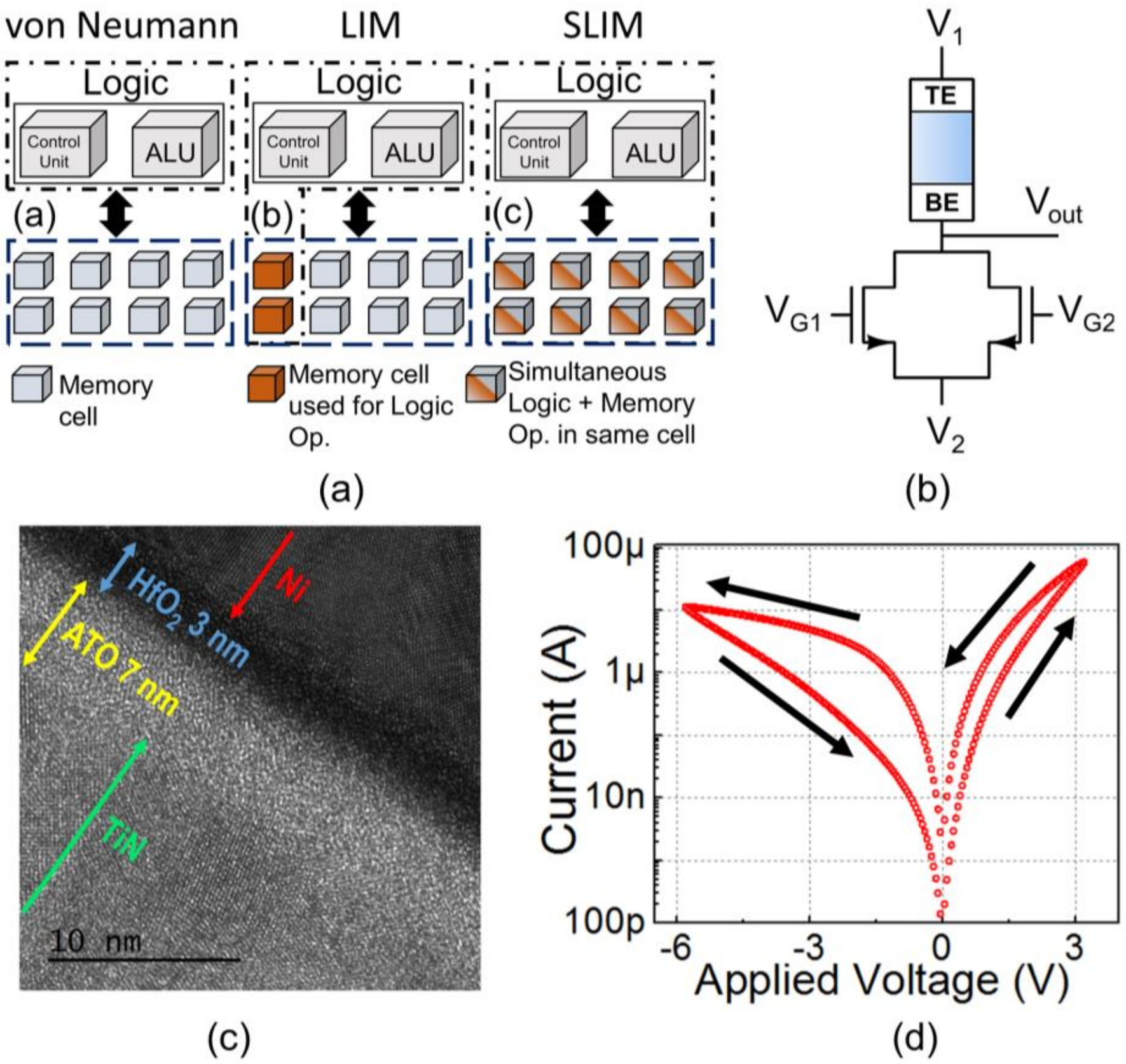

**Figure 1.** (a) von Neumann architecture with separate processing and memory units, LIM: Memory blocks can perform Logic, and SLIM: all Memory blocks capable to perform Storage and Logic operations simultaneously and non-destructively, (b) Proposed circuit schematic of 2T-1R SLIM bitcell, (c, d) HR-TEM and DC IV curve of bilayer Ni/$HfO_2$/ATO/TiN OxRAM device, fabricated for this study.

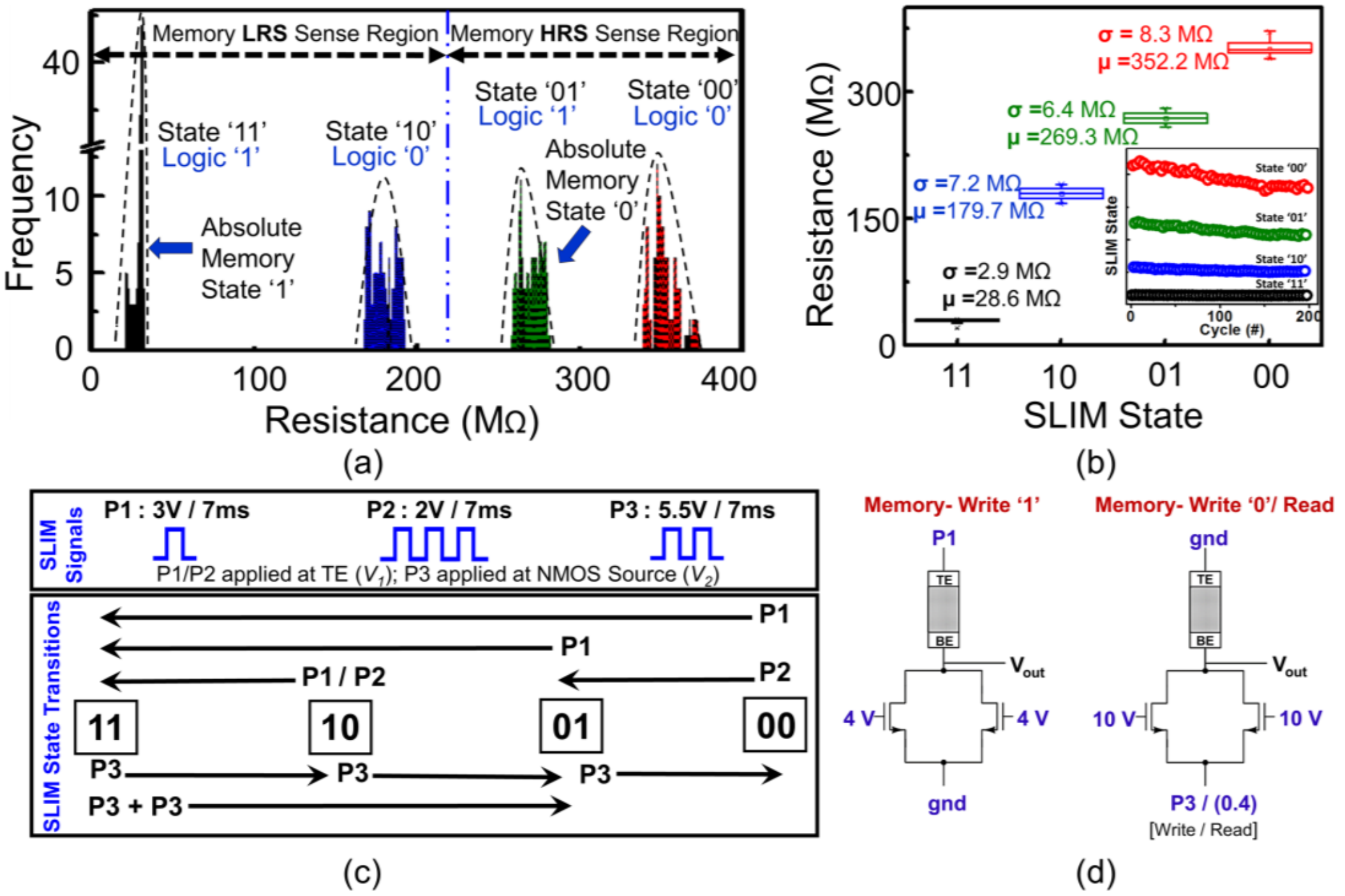

**Figure 2.** (a) SLIM state assignments for Logic and Memory operation. Note Memory LRS and HRS sense regions. Histograms show resistance distributions for 4 selected SLIM states on 2T-1R bitcell (>100 trials). (b) Resistance distribution for 4 selected SLIM states. Inset shows endurance for states: '11', '10', '01', '00' for 200 cycles. (c) Proposed SLIM programming signals and Memory-Logic state transitions for 2T-1R SLIM bitcell. All further SLIM operations use P1/ P2/ P3 pulses at $V_1$/$V_2$, and (d) Applied example signals for Memory Write '1' and Memory Write '0'. Read conditions are specified in brackets. [All pulse duration = 7 ms]

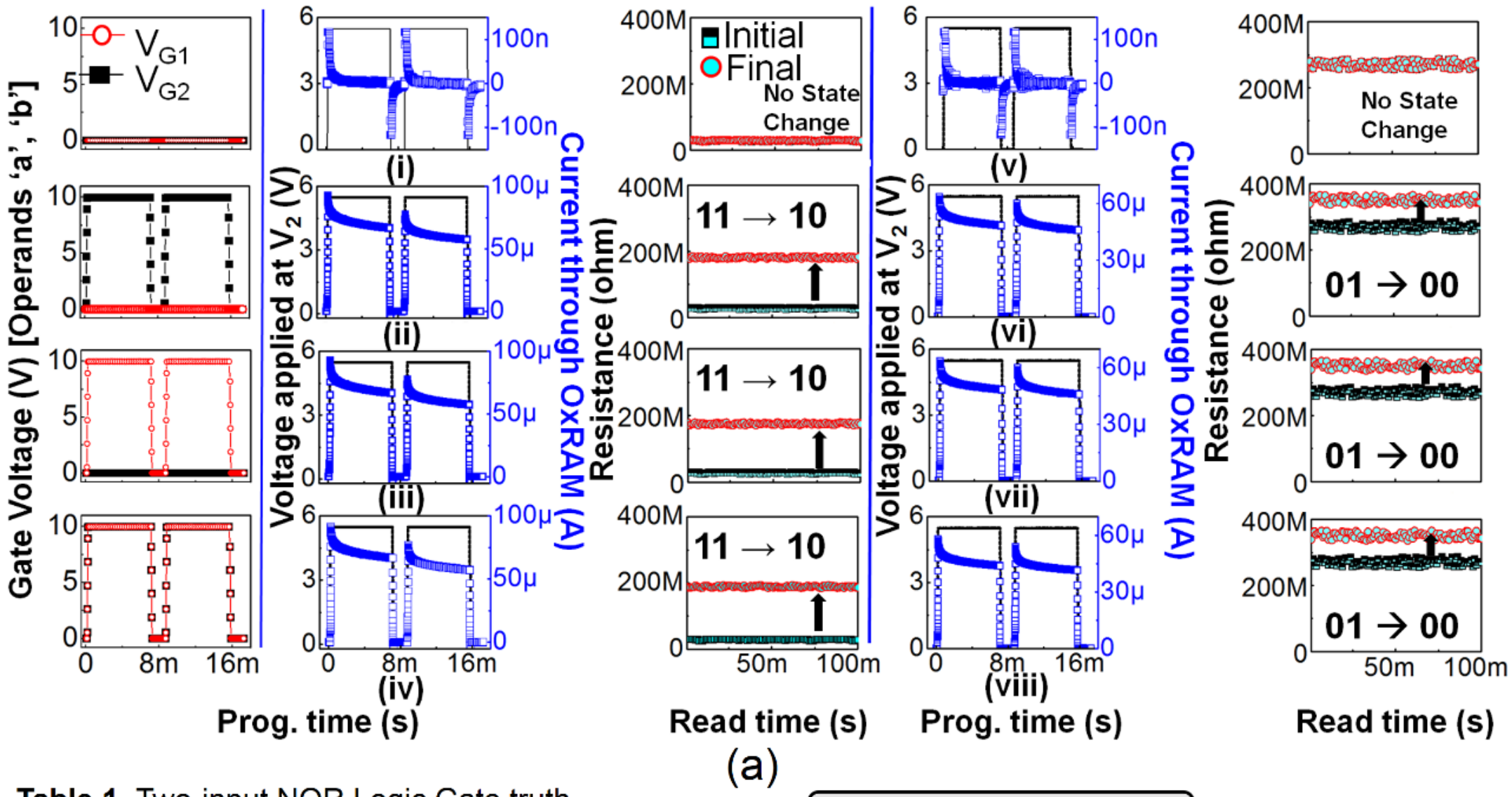

**Table 1.** Two-input NOR Logic Gate truth-table using our SLIM methodology

| Variables | | Stored Memory State: LRS | | | Stored Memory State: HRS | | |
|---|---|---|---|---|---|---|---|
| a | b | Device State Initial | Device State Final | Logic O/P | Device State Initial | Device State Final | Logic O/P |
| 0 | 0 | 11 | 11 | 1 | 01 | 01 | 1 |
| 0 | 1 | 11 | 10 | 0 | 01 | 00 | 0 |
| 1 | 0 | 11 | 10 | 0 | 01 | 00 | 0 |
| 1 | 1 | 11 | 10 | 0 | 01 | 00 | 0 |

State 11, 10: **Memory LRS**; State 01, 00: **Memory HRS**;
**Stored Memory State Preserved after all logic operations**

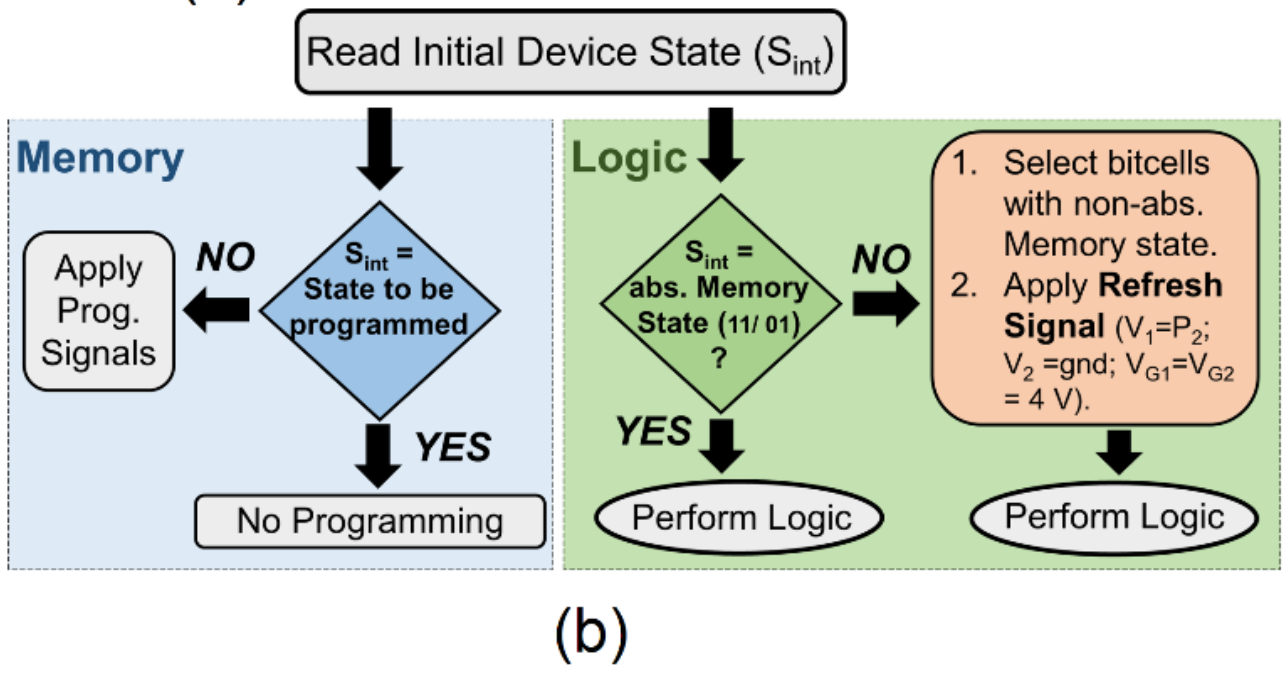

**Figure 3.** Experimental results for NOR logic implemented using 2T-1R SLIM bitcell with device's initial state: '11' (i-iv), and '01' (v-viii). Among the four operand combinations, OxRAM device switches to logic HRS state ('10' or '00') for a= '0', b='1'; a= '1', b= '0' and a=b= '1'. Blue: transient current through OxRAM device. Black line: P3 in all cases (applied signal). (b) Flowcharts for SLIM: Memory Write operation and Logic operations. Intelligent read is performed in both operations. Refresh scheme is an internal part of Logic operation.

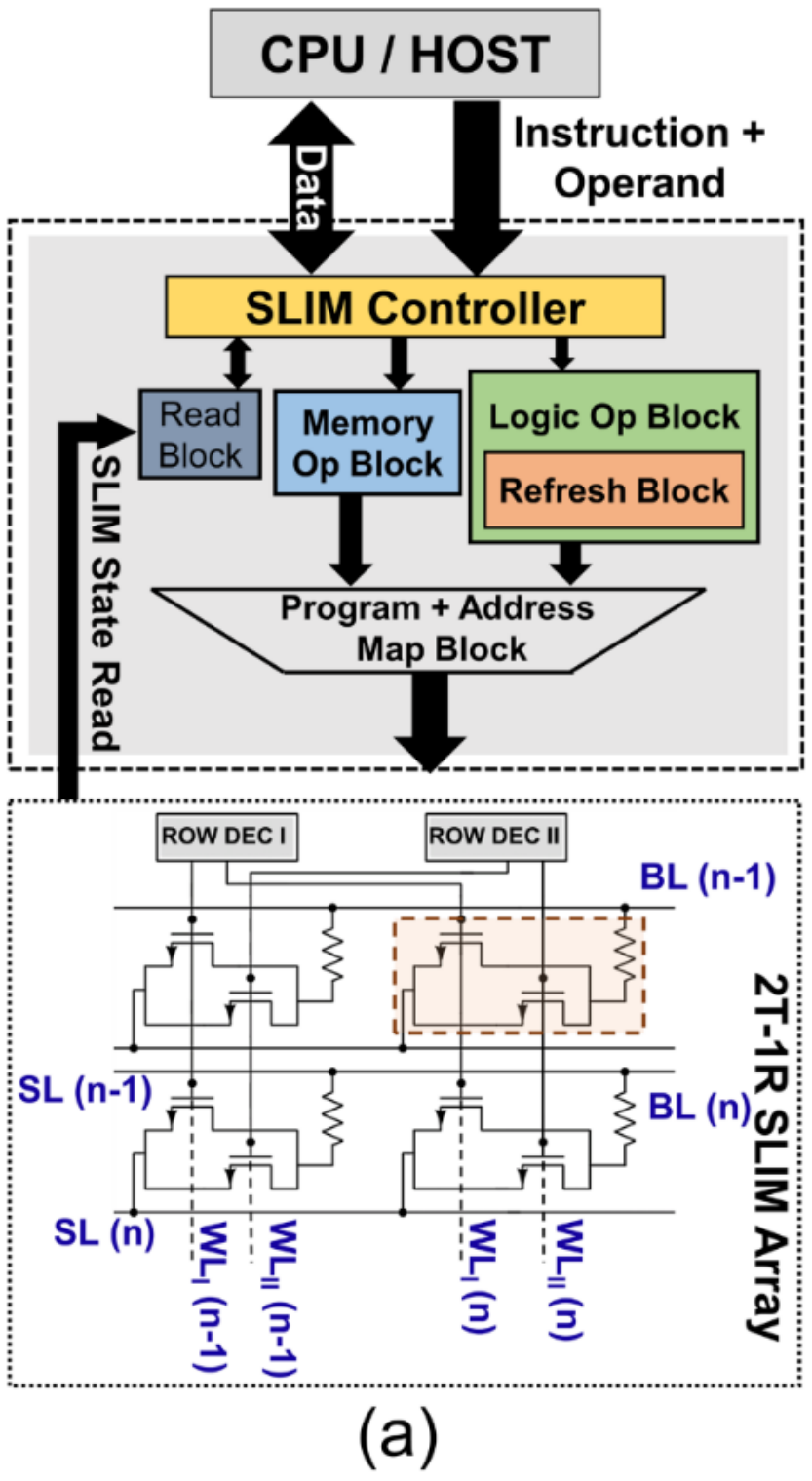

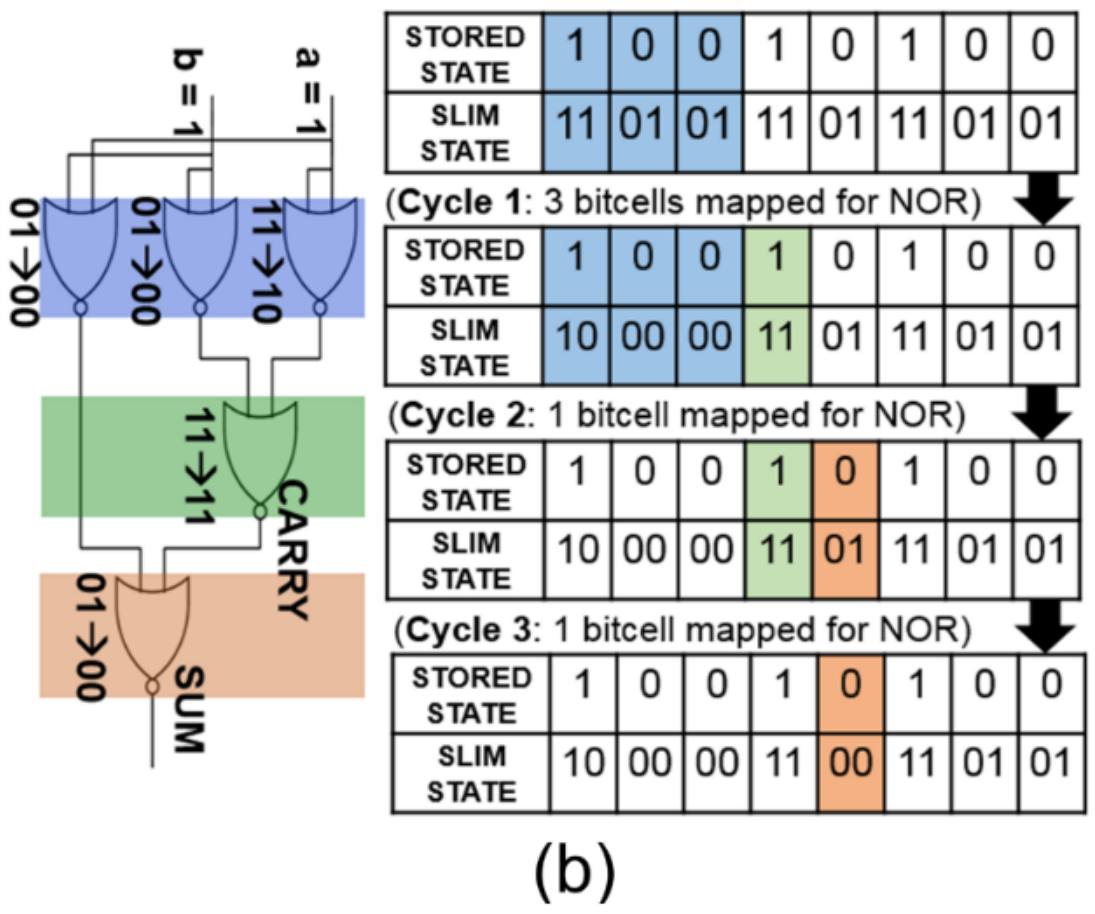

**Table 2.** Program scheme for other Boolean Logic operation using a SLIM bitcell.

| Logic | $V_1$ | $V_2$ | $V_{G1}$ | $V_{G2}$ |
|---|---|---|---|---|
| $\bar{a}$ | gnd | P3 | $a$ | $a$ |
| $\bar{b}$ | | P3 | $b$ | $b$ |
| $a+b$ | | $\bar{b}$ | $\bar{a}$ | $\bar{a}$ |
| $\overline{(a+b)}$ | | P3 | $a$ | $b$ |
| $a.b$ | | P3 | $\bar{a}$ | $\bar{b}$ |
| $\overline{(a.b)}$ | | $b$ | $a$ | $a$ |

**Figure 4.** (a) Block diagram of SLIM processing unit. Refresh block forms an internal part of the Logic operation block. User operands are passed first to the SLIM control unit. The control unit with other blocks maps Logic operations on the 2T-1R array.

(b) Logic cycle mapping of NOR based half adder using SLIM methodology. For generalization, arbitrary initial stored state is used.

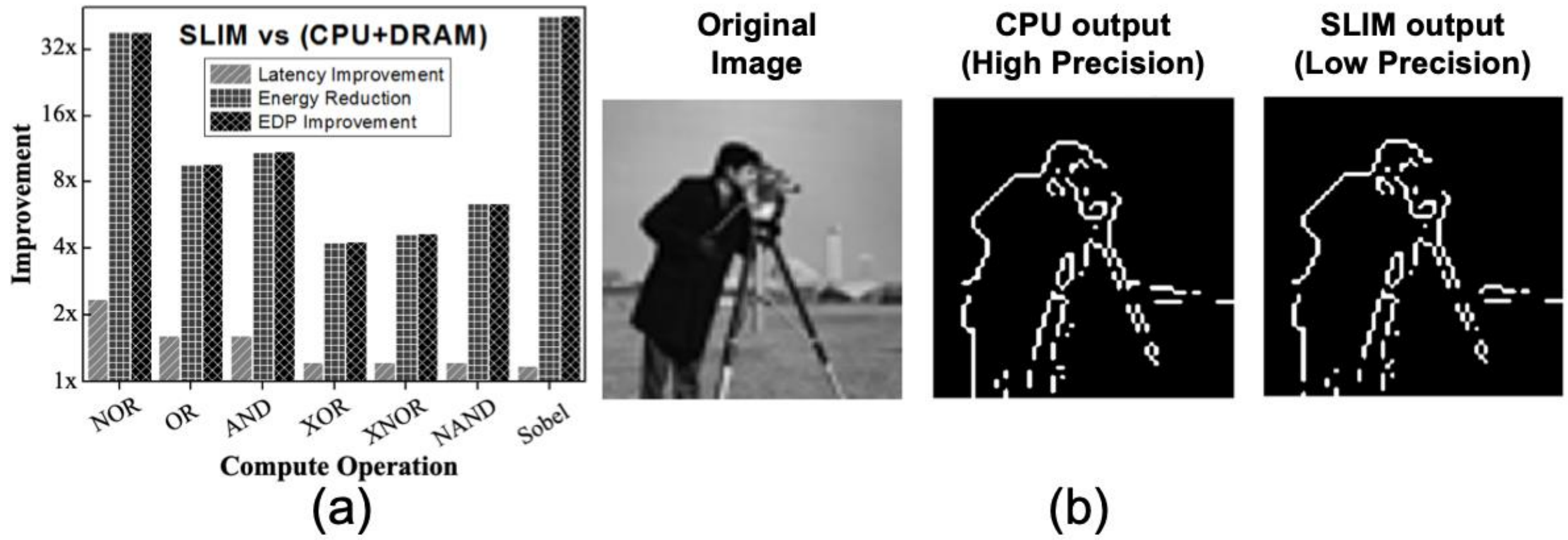

**Figure 5.** (a) Performance and Energy comparison for performing 64-bit Logic operations using SLIM bitcell array and conventional CPU architecture (fetching operands from DRAM (DDR3)), (b) Edge detection output of a (CPU+DRAM) and (CPU+ SLIM memory array) along with the original image.

**Table 3.** SLIM bitcell count, normalized Energy/Operation and normalized Latency count for different operations using SLIM NOR logic gate.

| Operation | #SLIM bitcells | Norm. Energy/Op | Norm. Latency |
|---|---|---|---|
| NOR | 1 | 1x | 1x |
| OR | 2 | 2x | 2x |
| NAND | 4 | 2x | 3x |
| AND | 3 | 1.75x | 2x |
| XOR | 5 | 3x | 3x |
| XNOR | 4 | 2.75x | 3x |
| 1-bit Half Adder | 5 | 3.37x | 4x |
| 1-bit Full Adder | 9 | 6x | 6x |

**Table 4.** Performance results for edge detection application using conventional and SLIM based system configuration[16, 17, 18, 19].

| System Configuration | Energy Delay Product (pJs) | | |
|---|---|---|---|
| | Data Transfer | Compute | Overall |
| CPU+DRAM | 1.31E-01 | 2.48E+05 | 2.48E+05 |
| SLIM | 1.68E-04 | 5.41E+03 | 5.41E+03 |
| Ratio | 783.44 | 45.89 | 45.89 |

**Table 5.** Comparison of different OxRAM based LIM methodologies proposed in literature w.r.t. current SLIM approach in this work.

| Methods | Ref. | Devices | Steps | Remarks |
|---|---|---|---|---|
| Sequential Logic | [10] | CRS: 1 | 3 | Retains only Logic output, Destructive Read |
| | [11] | BRS: 1 | 3 | Retains only Logic output, Needs rectifying behavior |
| | [12] | CRS: 1 | | |
| | [13] | BRS: 2 | | Retains only Logic output, Complex Integration |
| MAGIC | [14] | BRS: 3 | 2 | Retains only Logic output, Lack of signal restoration |
| 1T-1R | [15] | 1T-1R | 2 | Retains only Logic output |
| **SLIM (This work)*** | | 2T-1R | 1 | Retains Logic output + Initial Memory state *"simultaneously"* |

# Supplementary Information

## S1. Impact of C2C variability on DC IV curves for Bilayer OxRAM device

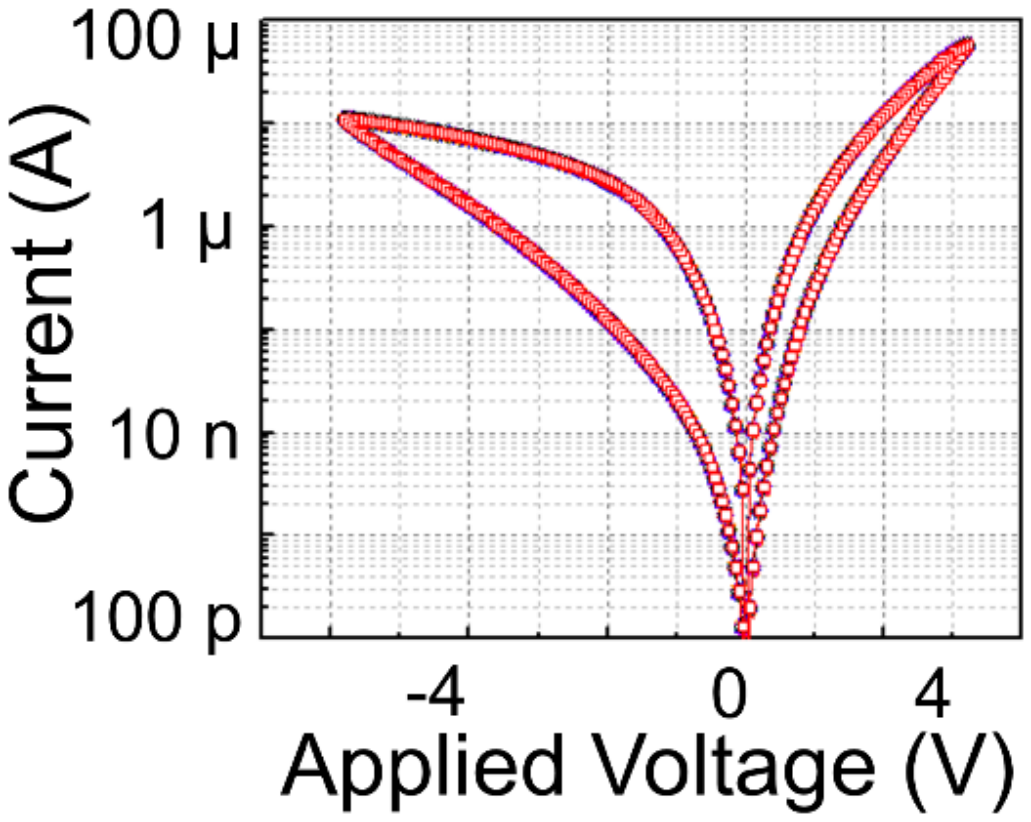

**Supplementary Figure 1.** DC IV curves of bilayer OxRAM device used in this study showing repeatable switching for 30 cycles. It is evident that device is quite stable over cycling.

## S2. Repeatable Analog conductance tuning characteristics for $V_{READ}$ = -0.4 V / -1.5 V.

For this analysis, a pulse train of identical SET pulses was applied, followed by a RESET pulse train, and resistance was measured at the read voltage ($V_{READ}$ = -0.4/ -1.5 V) after each particular pulse. Electrode's line resistance in the cross-bar array layout used is partially responsible for the D2D resistance variations observed.

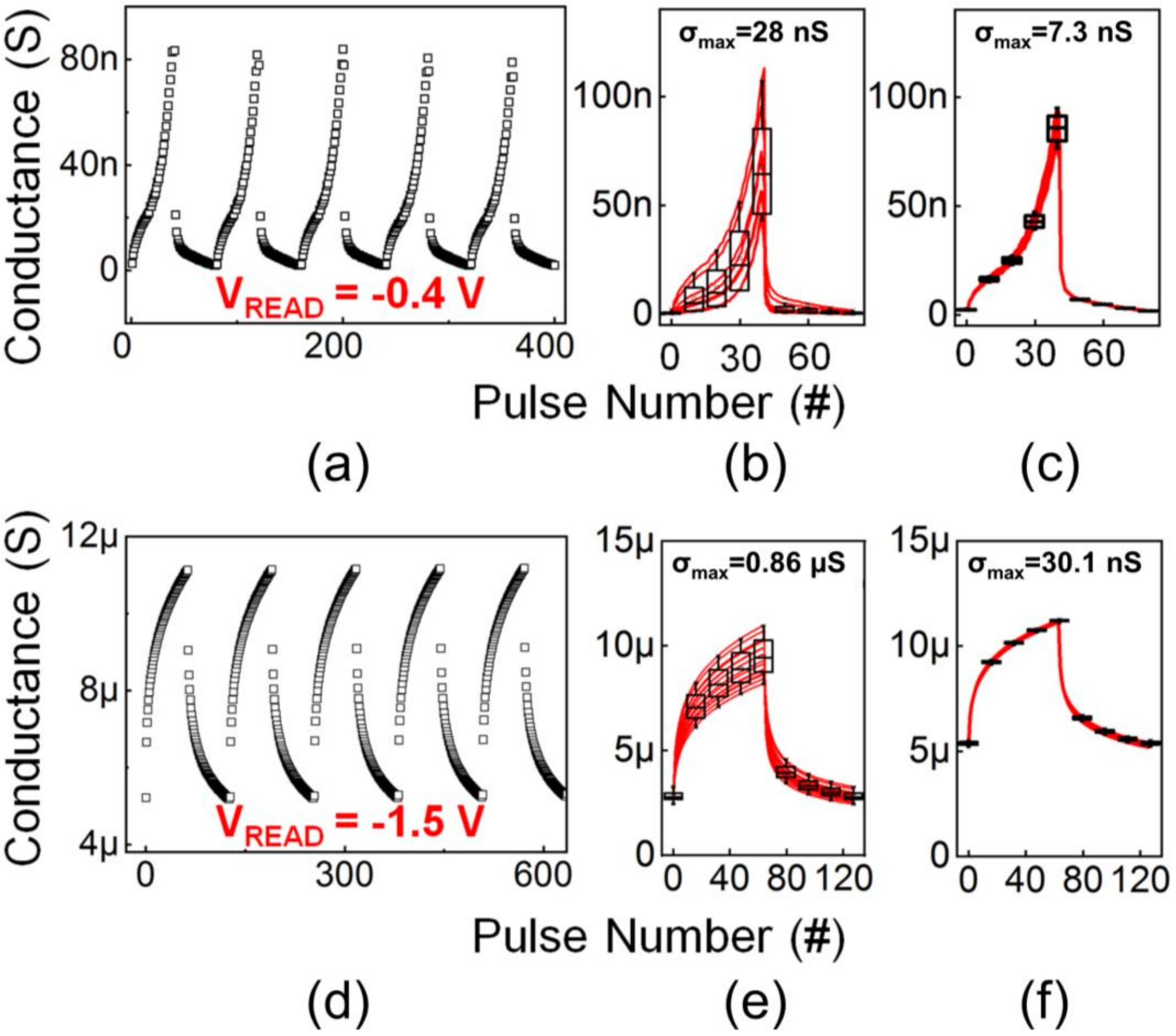

**Supplementary Figure 2.** Repeatable analog conductance tuning characteristics observed in the OxRAM device using identical SET and RESET pulses trains for (a) $V_{SET}$ = 3 V (10 ms) , $V_{RESET}$ = -5.5 V (10 ms) and $V_{READ}$ = -0.4 V. Effect of (b) D2D variability (from 10 devices), $\sigma_{max}$ = 28 nS (at mean= 57.3 nS), (c) C2C variability (for 30 cycles) , $\sigma_{max}$ = 7.35 nS (at mean= 79.6 nS ) has been observed. Similar characteristics for (d) $V_{SET}$ = 3 V (1 ms), $V_{RESET}$ = -3 V (5 ms) and $V_{READ}$ = -1.5 V, along with (e) D2D variability (from 16 devices), $\sigma_{max}$ = 0.86 µS (at mean= 9.49 µS) and (f) C2C variability (from 30 cycles), $\sigma_{max}$ = 0.031 µS (at mean= 11.2 µS) has been observed experimentally.

## S3. Transistor Characteristics

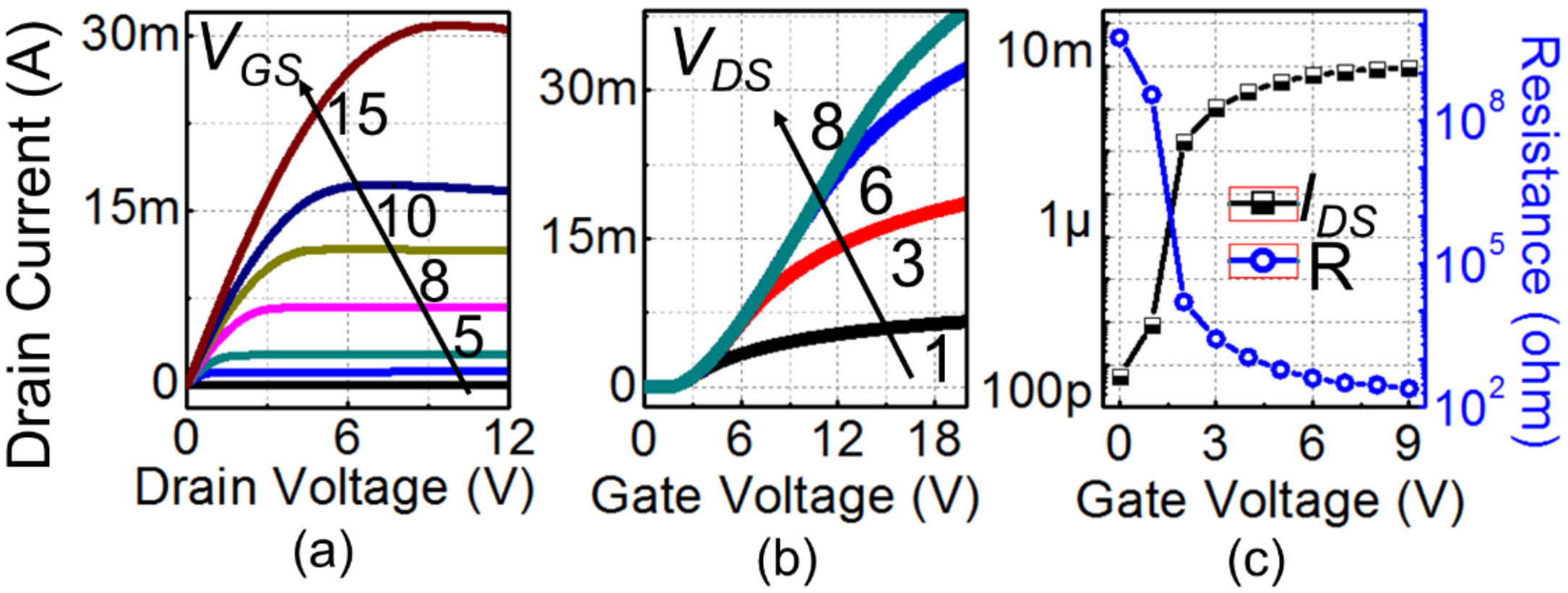

**Supplementary Figure 3.** Experimentally measured NMOS characteristics: $I_D$-$V_{DS}$ plot, $I_D$-$V_{GS}$ plot and relationship between NMOS transistor enforced compliance current ($Y_1$ axis) and ON resistance ($Y_2$ axis) with gate voltage ($V_{DS}$ = 3 V).

## S4. Experimental Setup used in this study

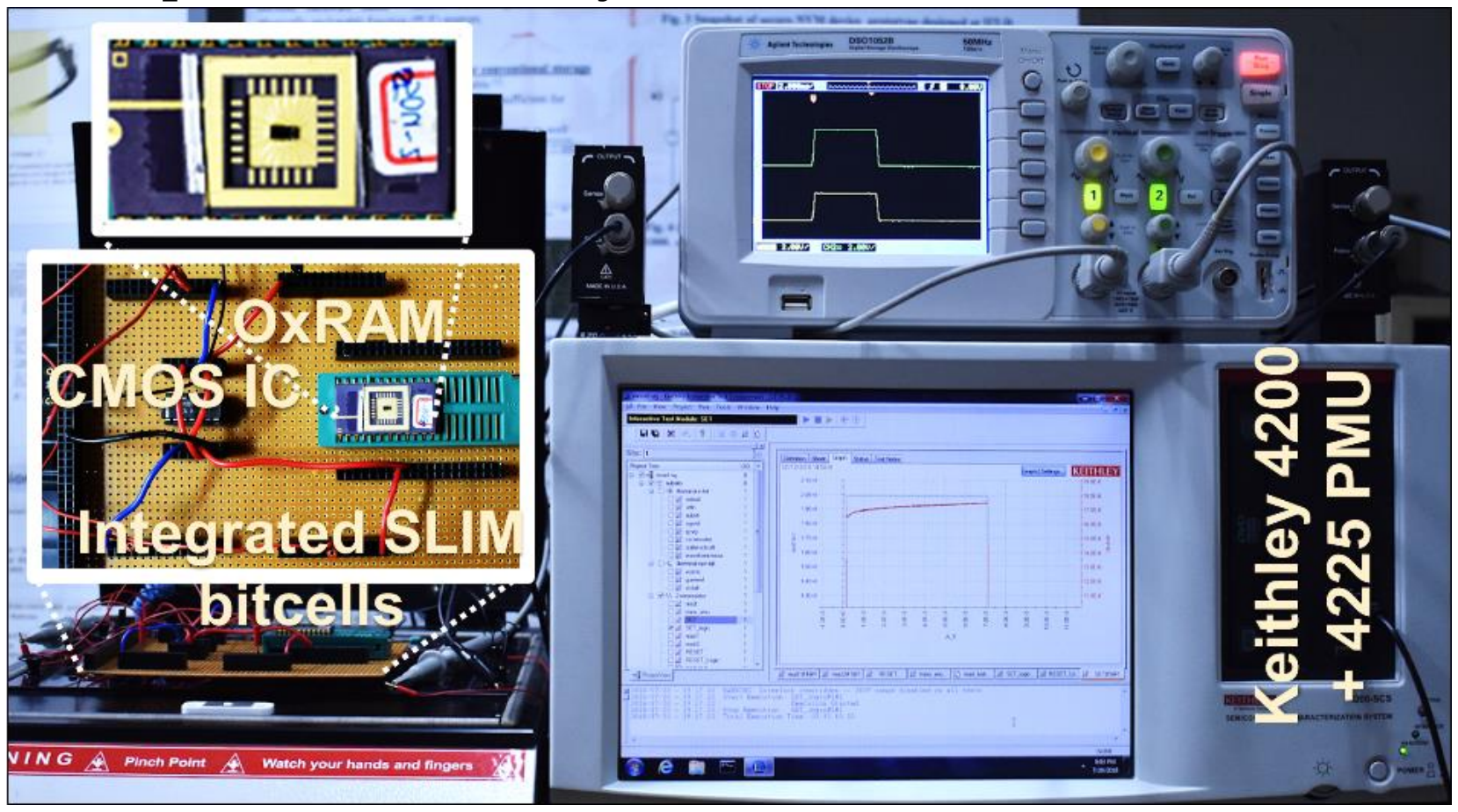

**Supplementary Figure 4.** Experimental setup used for SLIM characterization. It shows integrated 2T-1R SLIM bitcells, CMOS chip, OxRAM chip and parameter analyzer used for measurements.

## S5. Experimental results illustrating the Memory/Storage operation in SLIM bitcell

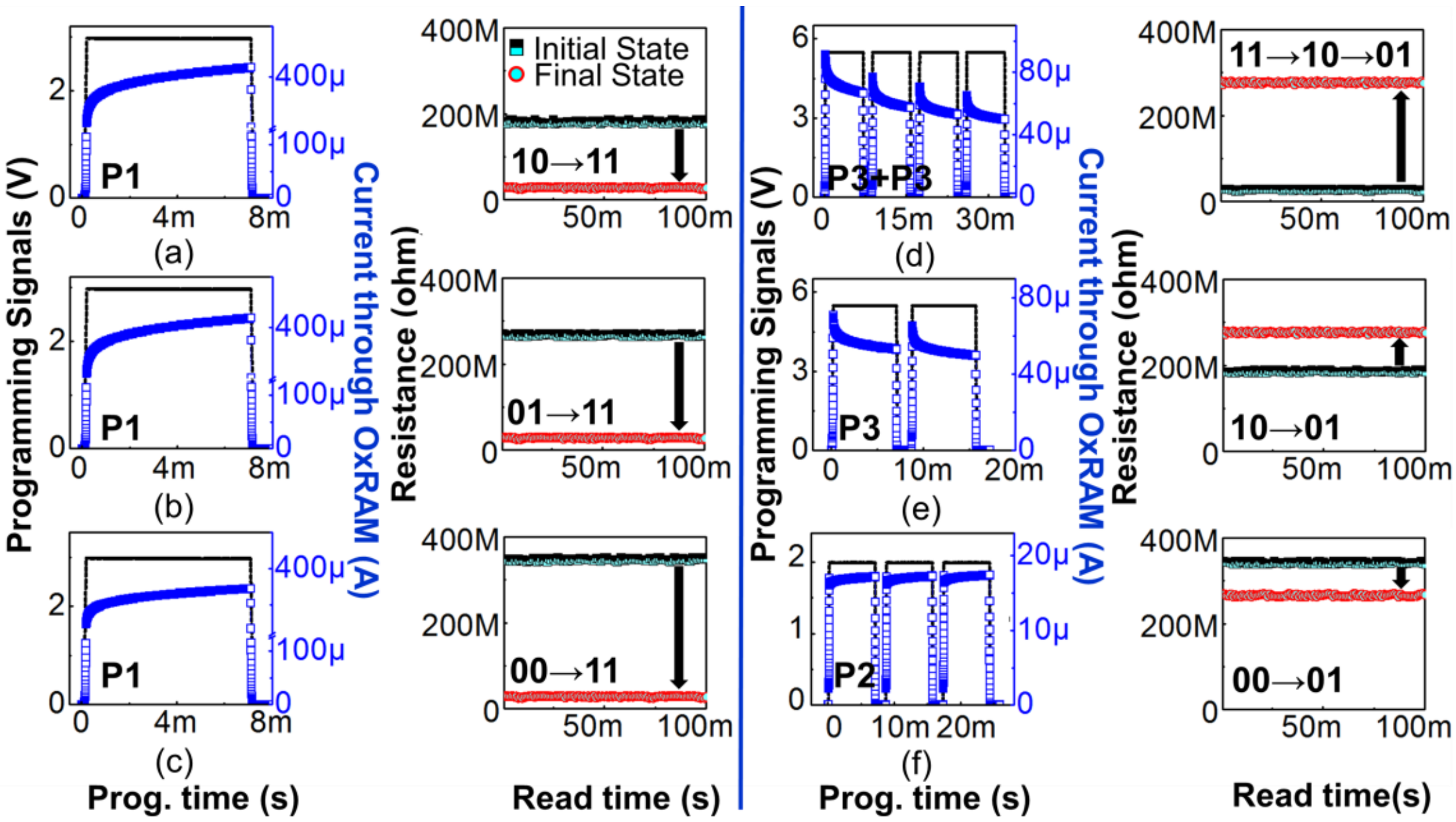

**Supplementary Figure 5.** Experimental measurements for Memory Write '1' operation (program device to state '11') with device's initial state as: (a) '10', (b) '01', and (c) '00'. Memory Write '0' operation (program device to state '01') with device's initial state as (d) '11', (e) '10' and (f) '00'. *Blue line:* transient current through OxRAM device. *Black line:* P1/P2/P3 (applied

signals). *Black square*: Initial resistance state, *Red circle*: Final resistance state post SLIM operation. Please note in (d, e), the transient current through OxRAM device falls due to gradual increase in non-volatile resistance with application of successive reset pulses.

## S6. Circuit schematics for different operand combination for realizing NOR Operation using 2T-1R bitcell

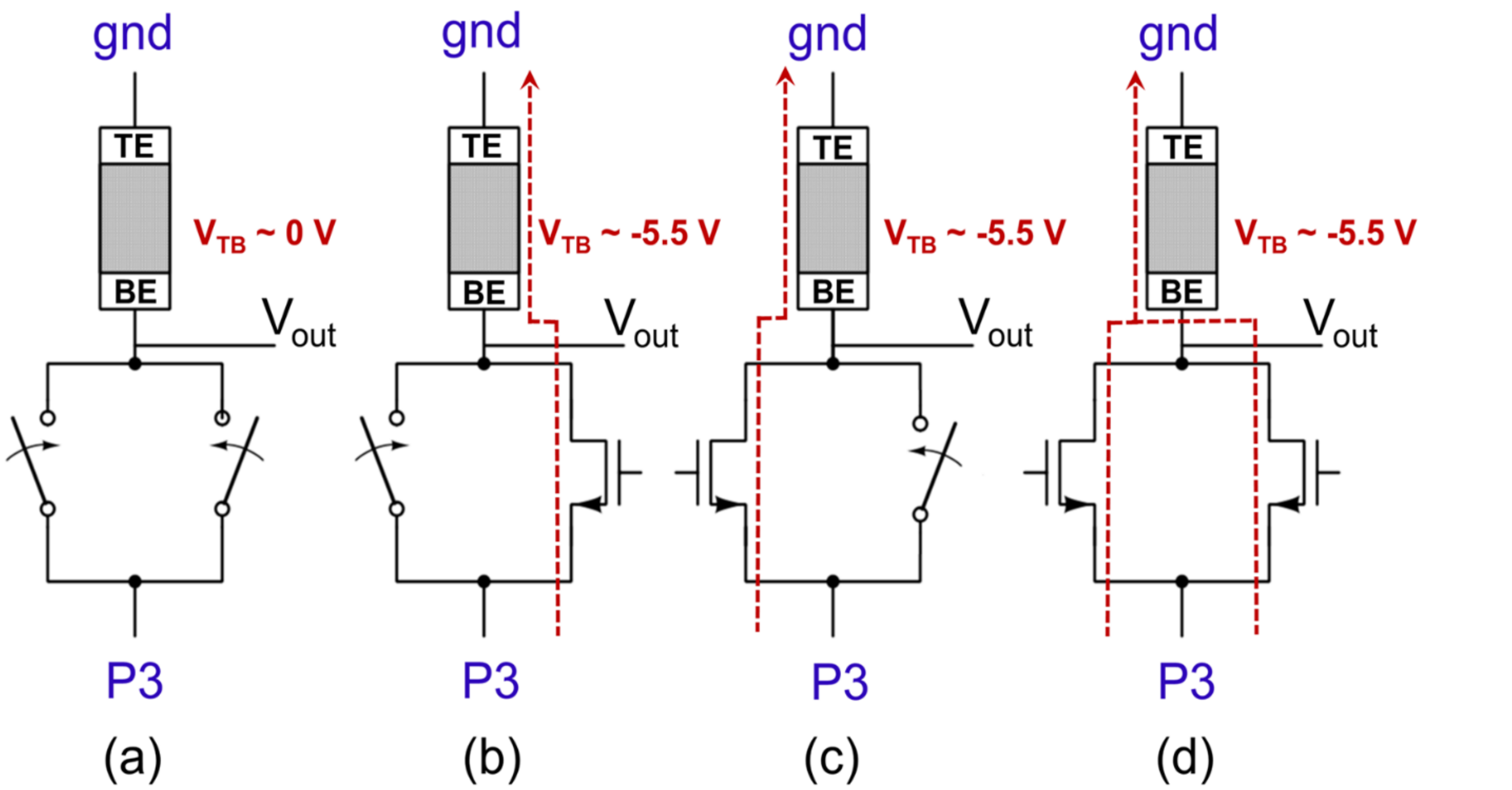

**Supplementary Figure 6.** Four possible input operand combinations (a) a = b = '0'; (b) a = '0', b = '1'; (c) a = '1', b = '0'; (d) a = b = '1'; corresponding to NOR truth table and proposed signal mapping for each case for the 2T-1R SLIM bitcell. [$V_{TB} = V_{TE} - V_{BE}$; $V_G$ = 10 V (7 ms long); $V_2$ = P3 = 5.5 V (7 ms long).

## S7. Principle of Refresh mechanism

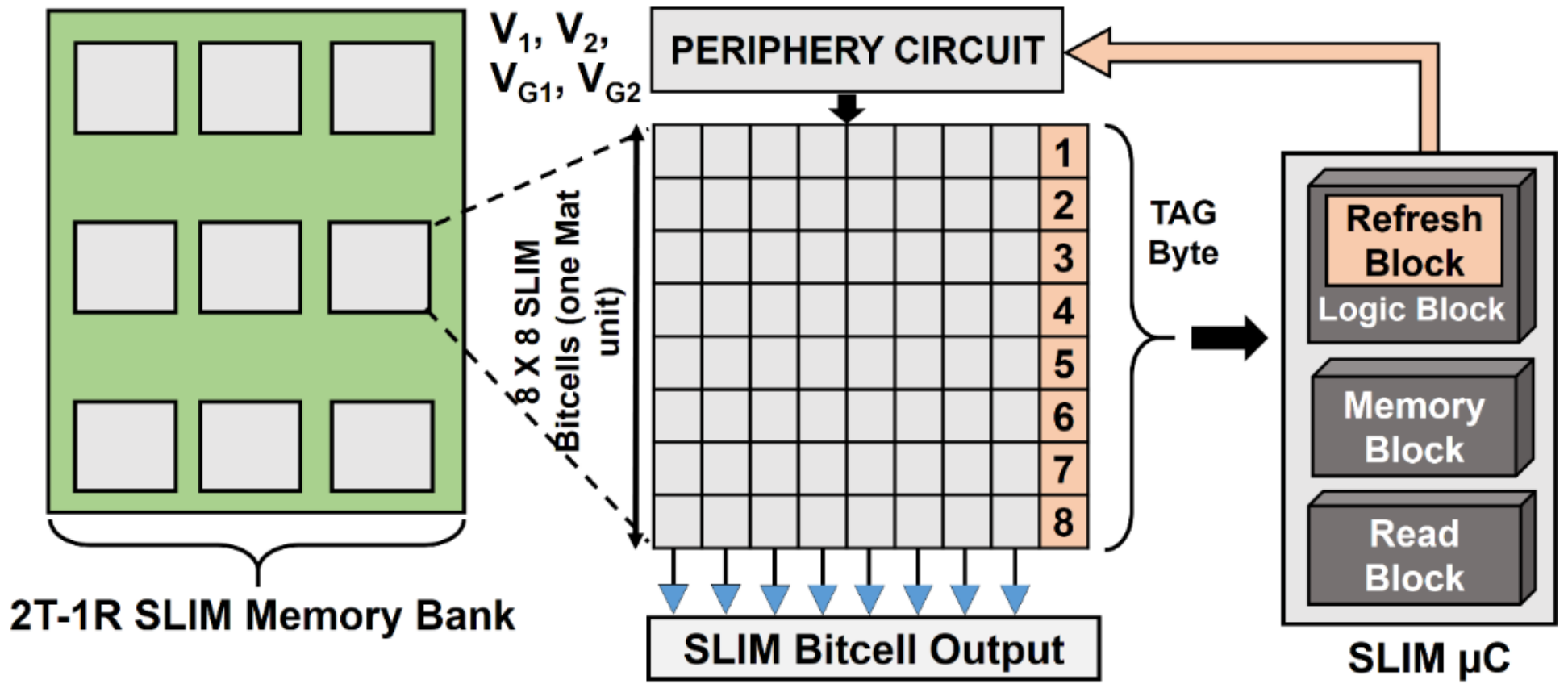

**Supplementary Figure 7.** Optimized Refresh Scheme corresponding to multiple SLIM mats. Each SLIM mat has 8x8 bits. 1 Tag byte is allocated to each SLIM mat to track previous operation (Logic/Memory). Within each Tag byte, single bit corresponds to 1 row of 8x8 SLIM mat. Tag Byte is initialized to zero once a fresh SLIM mat is used. Once a row is used for Logic operation, tag bit corresponding to it, is set high. When the tag byte contain all '1's, the Refresh block is triggered and it sends instruction to refresh the contents of complete SLIM mat. After refresh operation, all the SLIM bitcells in given SLIM mat will have absolute Memory states ('11'/ '01').

## S8. Advantage of using SLIM array in Memory Hierarchy

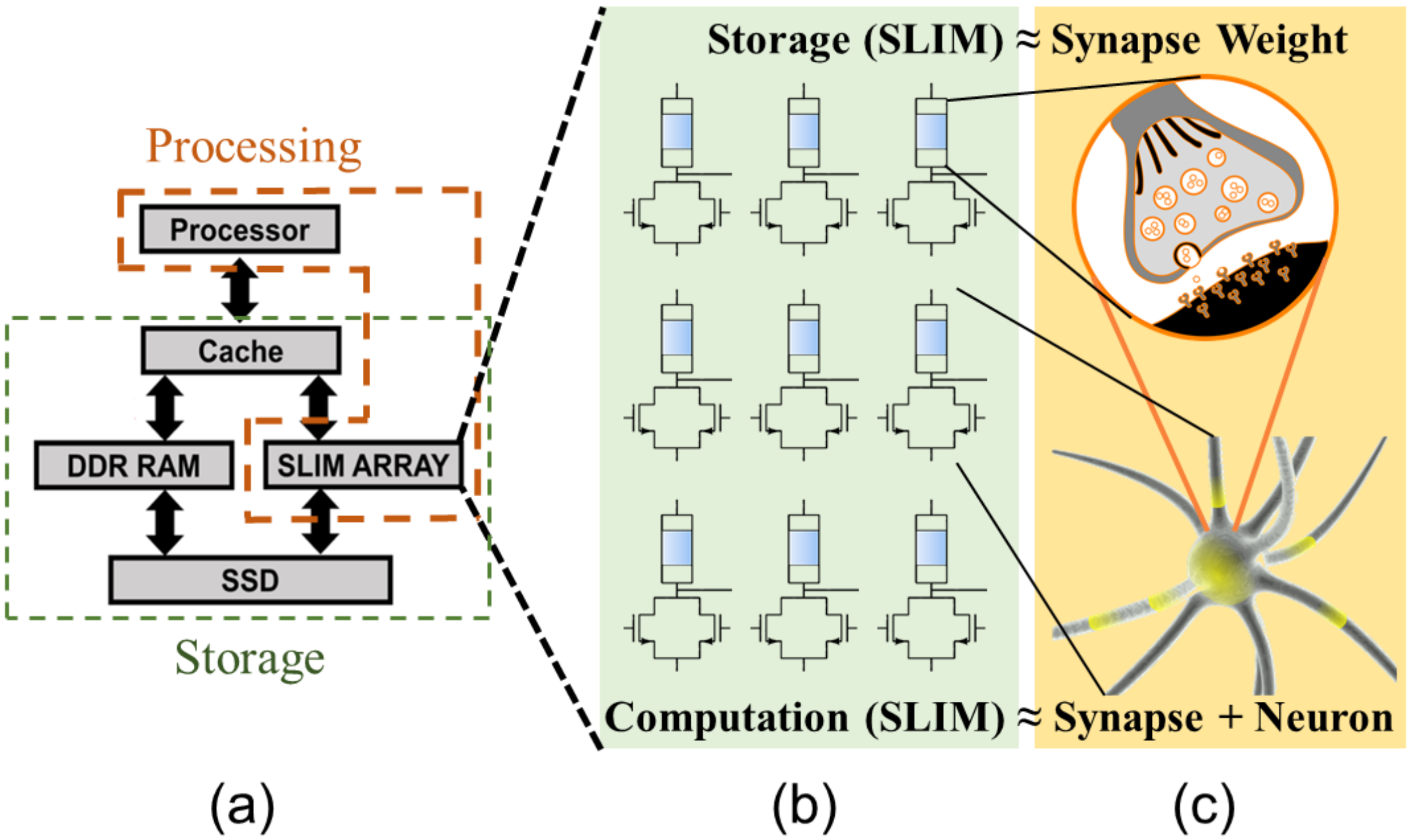

**Supplementary Figure 8.** (a) Memory Hierarchies: conventional (with DRAM) vs with SLIM array. SLIM array is capable of performing both Storage and Logic Operation. (b, c) Relationship between SLIM methodology and Neuromorphic computing.

## S9. Device parameters used for SLIM array

**Supplementary Table 1.** Device parameter used for SLIM array[16, 17].

| Parameter | Value |
|---|---|
| Switching energy | 10 pJ |
| Read energy | 0.25 pJ |
| Switching Latency | 10 ns |
| MAT configuration | 8x8 |
| MATs per bank | 32 |
| Total banks | 16 |